# All-optical manipulation of photonic membranes

Meisam Askari[1], Blair C. Kirkpatrick[1], Tomas Čižmár[2] and Andrea Di Falco[1*]

[1] SUPA, University of St Andrews, School of Physics and Astronomy, St Andrews, KY16 9SS, UK [2] Leibniz Institute of Photonic Technology, Jena, 07745, Germany

[2]Institute of Scientific Instruments of CAS, Královopolská 147, 612 64, Brno, Czech Republic

adf10@st-andrews.ac.uk

**Abstract:** Here we demonstrate the all-optical manipulation of polymeric membranes in microfluidic environments. The membranes are decorated with handles for their use in holographic optical tweezers setups. Our results show that due to their form factor the membranes present an 8-fold increase in their mechanical stability, respect to micrometric dielectric particles. This intrinsic superior stability is expected to improve profoundly a wide range of bio-photonic applications that rely on the optical manipulation of micrometric objects while studying single-molecule mechanics and sub-cellular scale transport processes.

## 1. Introduction

In 1970, A. Ashkin introduced the concept of Optical Trapping (OT) [1], demonstrating that a spherical bead suspended in a liquid can be trapped by a strongly focused Gaussian beam, due to the balance of scattering and gradient forces [2,3]. Today, OT is a versatile tool that is capable of exerting and measuring forces from fN to several nN [4], and is a mainstream technique adopted by a diverse range of applications from biological sciences [5–7] to atomic physics [8,9]. Using OT, it is possible to directly control the position of microscopic objects, without having to rely on diffusion or the flow of liquids in microfluidic channels. Instead, the trapped objects can be driven directly into the desired place and bioprocesses triggered on demand. Examples of biological applications of OT include the manipulation of single DNA molecules and experimental assays that have enabled insight into the behaviour of cells, organelles, and molecules [10–12]. In these examples, it is the molecules or other nano-objects are grafted onto the particle, which acts as a handle [13,14]. Since the biological specimens are not directly trapped inside the high-intensity regions of the trapping laser field, this approach also reduces possible photodamaging effects. The use of optical handles can be further extended to more complex shapes, to add functionalities to the optically manipulated object [15,16], from surface profilers [17], to steerable waveguides [18], microscopic pipettes [19], and inorganic nano-sheets [20]. The trapping dynamics in these cases are mostly dictated by the properties of the trapping beam (wavelength, power and numerical aperture), the refractive index contrast between the beads and the surrounding medium and the size of the beads.

OT mediated bio-experiments typically work by detecting the light scattered by the trapped particle (the handle) and correlating the extracted position with the forces acting on the particle, and thus on the grafted objects. To extract meaningful information on the biophysical processes of the studied objects, it is often required to monitor the position of the particle with extreme accuracy. For example, to visualize the stepping of a single enzyme moving along the DNA [21] it is required an OT able to resolve the position of the particle within ~1 Å.

In its simplest description, an optically trapped particle behaves like an overdamped mechanical oscillator, displaced out of its equilibrium by the Brownian Motion (BM) to which it is subjected [22,23]. This time dynamic is well described by the Langevin equation [24], which is a stochastic differential equation relating fast microscopic events (the collision events of the

particle with the surrounding fluid) to relatively slow macroscopic responses (the particle displacement):

$$m\frac{d^2x}{dt} + \gamma\frac{dx}{dt} + kx = \sqrt{2k_BT\gamma}\eta(t). \tag{1}$$

Here, $m$ denotes the particle mass, $\gamma$ is the viscous damping at the temperature T, $k$ is the trap stiffness, and $x$ is the particle displacement. The term $\sqrt{2k_BT\gamma}\eta(t)$ represents the stochastic Gaussian forces that drive the particle's dynamics. The viscous damping for a spherical particle can be calculated using Stokes' drag force and $\gamma = 6\pi\mu r$ for a particle with radius $r$ in a fluid with a viscosity of $\mu$. The uncertainty introduced by BM in the position of the particle is inversely dependent on the trap stiffness, which in turn is proportional to the power of the optical trapping laser [22]. Due to its spatial profile, a focused Gaussian beam typically produces a stronger trap in the plane transverse to its direction of propagation. Thus, the trapped particle position fluctuates more in the laser propagation direction, and the trap stiffness in this dimension is often the limiting factor for accurate particle tracking experiments.

A complete understanding of the stability of a trapped object must also include the effect of the various sources of noise that hamper the accurate measurement of its position in time. Since BM has zero average, to reduce its effect it is sufficient to track the position of the particle over long times (typically several tens of seconds), sacrificing the temporal resolution. However, for longer and longer acquisition times, external sources of noise such as the drift of the stage, and mechanical vibrations induced by air current become more relevant. Researchers go to extreme lengths to reduce the effect of these instabilities, from compensating the stage fluctuations to hermetically sealing the whole setup in inert gas, to isolating the whole laboratory from all kind of vibrations [21,25].

To date, all optical manipulation of Photonic Membranes (PM) has not been demonstrated and the advantages of PM as a grafting platform has not been tested. Here we demonstrate that by decorating the PM facet with handles placed at its edges, the PM can be manipulated programmatically, by means of established holographic optical trapping (HOT) techniques. Using high speed video acquisition technique and Principal Component Analysis (PCA) we studied the stability of the PMs and confirmed that the PM is intrinsically more stable than traditional spherical handles. It is believed that the form factor of the membrane introduces strong inertia to any movement in the direction perpendicular to its face (see the artistic sketch in Figure 1). In the following, we first present the fabrication procedure of the PMs and the experimental results and we then present the detailed analysis of their mechanical stability and discuss their suitability for bio-oriented experiments.

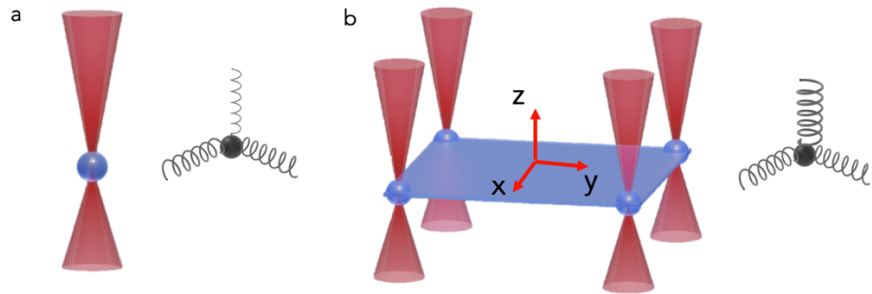

Figure 1 Artistic sketch of optical trapping. a) For spherical beads, the trap stiffness along the direction of propagation of the trapping beam is weaker than in the transversal direction. b) PMs presents intrinsic increased stability in the longitudinal direction in comparison to spherical beads.

## 2. Methods and Materials

To be able to demonstrate 3D trapping of membranes, bespoke membranes were fabricated using Electron Beam Lithography (EBL). These membranes were designed with integrated handles which made it possible to directly trap and manipulate high aspect ratio (15 μm x 15 μm membranes with a thickness of 90 nm) structure. A Holographic Optical Tweezer setup was designed and built to enable automatic manipulation of the membranes.

*2.1 Fabrication:*

The PM used in this experiment was fabricated in two stages. In the first stage membranes and alignment markers were fabricated and in the subsequent stage, handles were fabricated on top of the patterned structure. To fabricate the membrane structure, a silicon substrate was cleaned in using an ultrasonic bath and a sacrificial layer (Omnicoat) was spin-coated on the cleaned substrate at 1000 RPM for 60 seconds. The Omnicoat was soft-baked for 2 minutes at 230 °C. The photo-resist layer (Microchem SU-8 2000.5 Cyclopentenone 1:2) was spin-coated on the sacrificial layer at 5000 RPM for 60 seconds to produce 90 nm layer of photoresist on top of the sacrificial layer. The sample was soft-baked at 65 °C for 1 minute afterwards the temperature increased to 100 °C and soft-baking continued for further for 4 minutes. The prepared sample was patterned using EBL (Raith eLINE Plus system). We then performed a post-exposure baking step at 100 °C and developed the samples by immersion in the Ethyl Lactate (EC) solvent. The handles were defined by spinning a 1.5 μm layer SU-8, with identical exposure and development procedure. A typical exposure run allows the definition of several 100s of thousands of membranes. MF-319 (MicroChem) was used to remove the sacrificial layer of Omnicoat and release the PM structures. By adding deionized water and gradually extracting the liquid from the top layer of the vial, the concentration of MF-319 was reduced and replaced by deionized water. After lift-off, the water solution containing the membranes can be placed in a standard microfluidic chamber for the experiments.

*2.2 Experimental setup:*

We used a standard Holographical Optical Tweezers (HOTs) system, consisting of a LuxX compact CW diode laser, (Omicron Laserage Laserprodukte GmbH) operating at 830 nm with a maximum output power of 230 mW. A spatial light modulator (Boulder Nonlinear Systems 512x512), has been integrated into an inverted microscope layout with a water-immersion, high-numerical-aperture objective lens (Olympus UPLSAPO 60XW). Control of the HOTs is granted through a custom-built LabView virtual instrument, providing a graphical user interface for the end-user. Trap analysis was done using a high-speed video acquisition technique. Basler piA640-210gm was used in this experiment. By restricting the sensor area to an array of 100 by 100 pixels the video acquisition rate was increased to 1000 frames per second.

## 3. Results

Photonic membranes are built through a multi-step fabrication process, following a consolidated approach to make flexible metasurfaces on polymeric substrate [26], based on electron beam lithography. The membranes and handles are defined on a rigid carrier pre-coated with a sacrificial layer, using a negative tone electron beam resists, with thickness ranging from a few tens of nanometres to several microns. Most importantly, this approach can be extended easily to pattern the surface of the membranes with polymeric, dielectric or metallic features at the nanometers scales, to define metasurfaces with bespoke photonic functions. The yield of the process is very high and a typical sample with an area of 1cm$^2$ contains in excess of 100K PMs. Figure 2(a) shows a scanning electron microscope image of a uniform PM with square handles and Figure 2 (b) shows PMs of different scales and with different motifs built into them, after e-beam exposure and development. The PM is subsequently released chemically from the substrates and introduced into a microfluidic cell where they can be manipulated via a traditional HOT setup (see e.g. ref [27]). Using HOT, a trap per corner is generated, and the PMs are then manipulated by addressing the four traps individually. We demonstrate this control by moving the PM through the microfluidic environment in a fully automated sequence of steps, as shown in the supplementary video.

Keyframes from this sequence are shown in Figure 2(c)-(f). To aid the eye in distinguishing PM orientation, a one-dimensional grating was added to its surface.

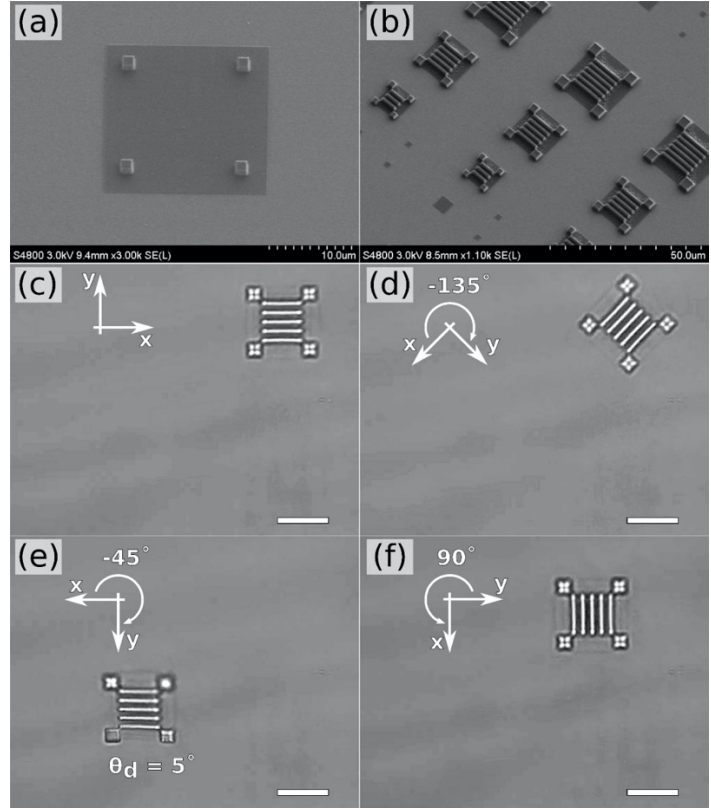

Figure 2 Photonic metasurface fabricated using e-beam lithography and manipulated by trapping the handle structures patterned at the four corners. (a) SEM picture of the uniform membrane, (b) array of PM with a 1D grating pattern. Panel. (c)-(d) Bright-field microscope image of isolated particle and the PM with particles adhered to PM (c) shows a PM in the start position, trapped via one optical trap per handle (traps not visible). (d) The PM is rotated clockwise about the optical axis of the trapping beam through an angle of 135°. (e) The PM is then translated through the sample and tilted out of the plane by approximately 5°. Held in this orientation, it is then rotated about the optical axis through a further 45°. (f) Finally, the PM is brought back parallel to the XY plane and then rotated through 90° while being translated back towards its starting position. The scale bar is 10 μm.

To characterise the optomechanical dynamics of the PM, we used a plain, untextured membrane with a side of 15 μm and square handles with sides of 2 μm. To track its position in time, we fixed to its surface a bead of 1 μm diameter. This approach guarantees a straightforward quantitative assessment of the motion of the membrane, which relies on the well predictable scattering of the spherical bead. In the following, when referring to the dynamics of the membrane we will implicitly assume that we are indeed observing and describing the dynamics of the bead attached to it. The positions of the isolated bead and of that of the PM were tracked in three dimensions, using video acquisition with a CMOS camera, and a custom-built algorithm, based on the shift property of Fourier transform and PCA [28]. A typical reconstructed trace of the position in space vs time is shown in figure 3a), relative to a trapping laser power of 4mW, measured directly after the microscope objective. In the case of the membrane, the power was distributed equally across the 4 handles, which were individually trapped with 1.25 mW each. It is particularly instructive to create a scatter plot of these trajectories, as shown in figure 3b) for both the bead and the membrane. To avoid crowding the figure, we selected a 2s period of the traces, greyed out in panel a).

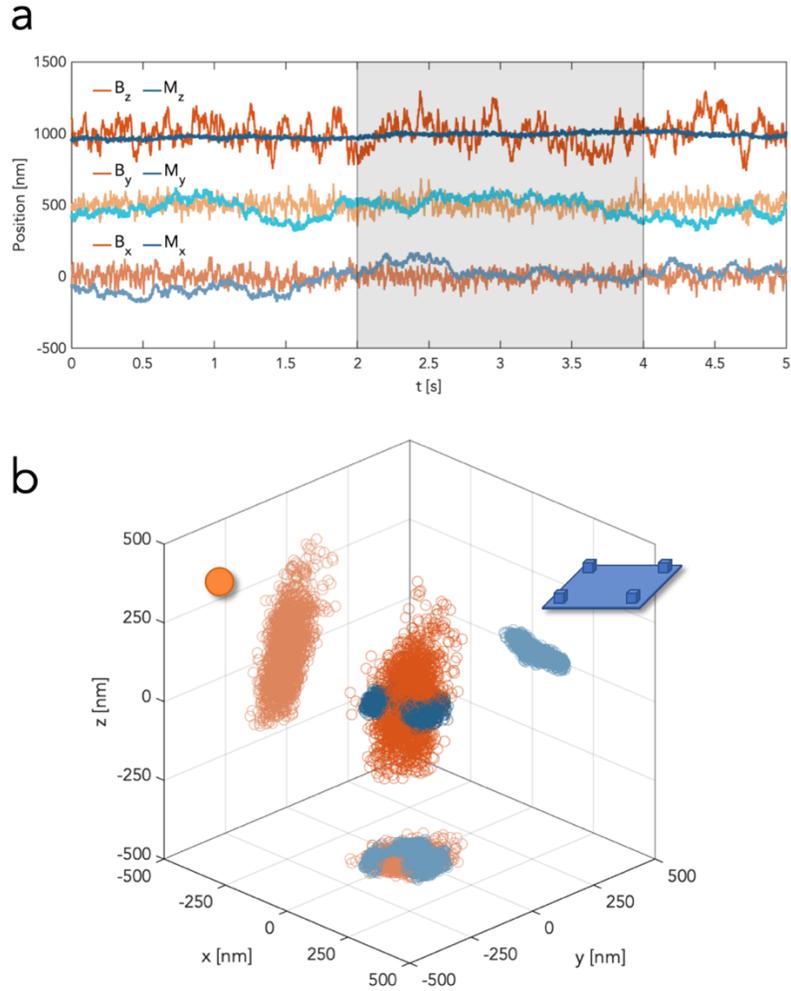

Figure 3 The stability of trapped objects is dependent on their form factor. Shown here are particle. (a) membrane trajectory in X, Y and Z directions and (b) 3D reconstructions of the BM undergone by a particle in orange and membrane in blue. It is immediately noticeable that the BM along the z-axis for the case of the flying carpet is greatly suppressed.

### 4. Discussions

In the following analysis, we will assume that the motion of the membrane is well described by the standard formulation of the Langevin equation (1). However, since the dynamics are dictated by the form factor of the PM, the hydrodynamic drag coefficient of a sphere cannot be used to describe the motion of the fiducial bead attached to the PM. Nonetheless, the typical figures of merit that describe the motion of an optically trapped object can be extracted from the trajectories shown in Figure 4a). To this extent, for simplicity sake, here we assume that the motion of the membrane is substantially unaffected by the bead attached to it. Additionally, the membrane is assumed to be not deformable and only subject to rigid translations. Both of these assumptions do not affect the validity of the results. A qualitative investigation of Figure 4a) suggests readily that the PM is more subject to low frequency and relatively large departures from its equilibrium position respect to an isolated bead. On the other hand, high-frequency oscillations appear to be substantially suppressed. This can be better appreciated in the scatter plots of the positions of both an isolated bead and a PM shown in Figure 4b), which corresponds to a 7.5 fold reduction of the standard deviation of the dynamics in the direction perpendicular to the direction of propagation of the trapping beam. It should be noted that for higher and higher trapping power this *advantage* reduces (doubling the power of the trapping beam reduces this factor to ~5x). However, in most biological assays it is particularly important to keep the optical power at its minimum, to avoid depositing large optical energies in close proximity to the analyte of interest. This initial consideration encapsulates the most striking difference between these two optically trapped objects and motivates the rationale for considering the PM as a valid platform in optical trapping bio-assays.

A more quantitative analysis can be completed extracting the power spectral density (PSD) from the trajectories of Figure 4, which are described by the analytical formula $P(f) = \frac{k_B T}{2\pi^2 \gamma} \frac{1}{(f^2 + f_c^2)}$. Here, $f_c = (k/2\pi\gamma)$ is the corner frequency, which is given by the ratio between the spring constant $k$ (proportional to the trap stiffness) and the friction coefficient $\gamma$. From a physical point of view, for frequencies below $f_c$, the motion of the object is determined by the trap potential. For frequencies higher than $f_c$ the dynamics is diffusive and driven by the BM. Remembering that the Einstein diffusion coefficient is $D = k_B T/\gamma$, when plotted in log-log scale the PSD shows a plateau for low frequencies at a value $D/(2\pi^2 f_c^2)$ and a high-frequency behaviour dictated by the term $D/(2\pi^2 f^2)$.

For the case of the isolated bead, the tracking along the longitudinal direction was calibrated assuming the that diffusion constant of the bead is the same in all directions. A fit of the PSD shown as dashed lines in Figure 2 gives a diffusion constant D=0.48 μm²/Hz and a trap stiffness of $k_{x,y}$=2 fN/nm and $k_z$=0.41 fN/nm, in the transverse and longitudinal direction respectively. These values are in line with what expected by a HOT system, given the low power and the fact that we didn't adopt any elaborate strategy to increase the stability of the system. In the Supplementary Information we show that $k_x$ increases linearly with the trapping power, as expected for well-behaved setups. The PSD of the membrane trajectories does not show obvious trapped dynamics, despite the evidence that PMs can be manipulated effectively in the medium (as witnessed by the video in the SI). This is mostly due to reaching the noise floor of our system. Despite this shortcoming, the analysis of the PSD can be used to extract the diffusion coefficients for the PM, relying on the calibration of the tracking procedure completed for the isolated bead case. Due to the form factor of the PM, these vary from a value of $D_{xy}$=0.01 μm²/Hz to $D_z$=1e⁻⁴ μm²/Hz, for the transverse and longitudinal directions, respectively. These values are obtained from the fit of the $f^2$ slopes of the PSD. Most remarkably, already at a frequency of ~ 10Hz, the dynamics of the membrane is so stable that goes below the detection limit of the system, as witnessed by the flattening of the PSD in the z-direction of the PM, shown in the third panel of Figure 4a.

To evaluate the usability of the PM as a stable macro-handle for bio-photonics experiments, it is useful to use the trajectories in Figure 3 to calculate the Mean Square Displacement (MSD) as a function of the time interval $\tau$, which gives a measure of the displacement within a given lag time of the trapped object from its equilibrium position. Due to the action of the trapping potential, a trapped object will remain in proximity of its original position, with an average distance dictated by the trapped stiffness. Figure 4b shows that after an initial period proportional to $2D\tau$, in which the trapped objects diffuse due to the BM, they reach a plateau, with value inversely proportional to the trap stiffness. In Figure 4b, we marked these values for the isolated beads, using the same parameters extracted by the fit of Figure 4a, without any additional adjustment, showing an excellent agreement between the different methods of analysis. From the MSD of the membrane, it is clear that the PM has a somewhat less straightforward behaviour. The dynamics along the transverse direction only plateaus around 1 s, at an intermediate trap stiffness between $k_{x,y}$ and $k_z$ of the bead. In comparison, the PM appears to be more stable along the $z$ direction, with an MSD roughly 40 times lower than that of the isolated bead. The comprehensive analysis of the trajectories, PSD and MSD of the motion of the PM compared to that of an isolated bead provides a clear insight into the different type of dynamics that governs them. Whereas an isolated bead moves very quickly in all directions and needs to be trapped with highly focused and relatively high power beams to be stabilized, a PM tends to move very slowly and is very stable even at extremely low pumping power. However, this also means that while for a bead it is possible to average out the noise simply by acquiring its trajectory over a longer and longer periods, a PM is more exposed to the effects of slow and long terms noise factors. Therefore, PM appears to be particularly promising for experiments that require very short acquisition times at low trapping power [29], [30].

Finally, PMs introduce a variety of appealing features in an optical trapping setup. Their extended surface can be patterned to create optical two-dimensional metamaterials, also known as metasurfaces (MSs). MSs are the most versatile photonic devices to date that break dependence of optical elements on the propagation effect by engineering the interaction of light with designed optical antennas placed on thin films [31]. The working principle of the metasurfaces is based on changing the shape, size and orientation of antennas placed with subwavelength gaps to produce a spatially varying optical response. A vast range of application such as thin lenses, filters and polarizes are envisioned for metasurfaces [31]. Metasurfaces can also be engineered to create attractive or repulsive forces, depending on the polarization of the light used to probe them [32], and it's been argued that they can even be used as solar sails [33]. All-optical manipulation of PM can also be used with purpose-built metasurfaces that can selectively interact with cells and biological materials in the microfluidic channel. Light-sheet microscopy, embedding microlasers in the membrane and producing microlenses are some of the notable applications of using metasurfaces in the microfluidic channel.

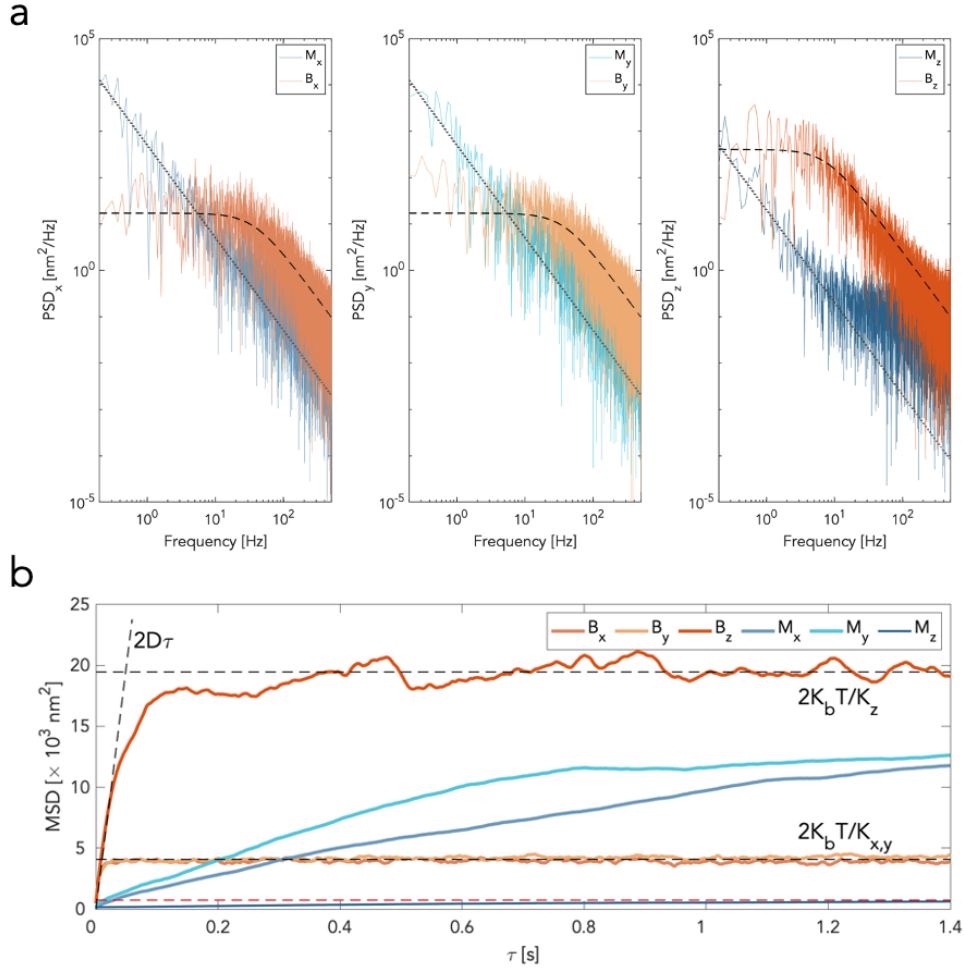

Figure 4 (a) shows the power spectral density (PSD signal of the isolated bead in blue and the trapped PM in orange for different trajectories. The dashed black line is the Lorentzian fit of the PSD signal that signifies the corner frequency. (b) demonstrate the MSD signal for the trapped membrane and particle. After the initial diffusion period, the MSD graph for the trapped particle in Z plateaus at a much higher value compared to the same graph for the membrane.

## 5. Conclusion

In conclusion, PM structures fabricated using EBL were suspended in water and manipulated using HOT. We demonstrated that by designing PM structures with appropriate handles, it is possible to achieve 6 degrees of freedom in the manipulation of the PM in the microfluidic chamber. Using PSD and MSD methods, we analysed the trapping stability of the spherical particle and PM structure in the trap and we reported that the PM stability in the laser propagation direction is far superior to the trapped particle. Using PCA we extracted the particle and PM trajectory in space, and we established that in the laser propagation direction, the PM stability is almost 8 times better than that of the particle, which can solve the inherent issue with trap stability in the laser propagation direction. Using PSD, we examined the PM behaviour at different frequencies. The trapped PM damps high-frequency vibrations produced by BM. Therefore, by utilizing high trap stiffness and PM, it is possible to damp both low frequency and high-frequency vibrations.

**Funding.** The project was supported by the European Research Council (ERC) under the European Union Horizon 2020 research and innovation program (Grant Agreement No. 819346).

**Acknowledgements:** BC fabricated the sample and collected the experimental data. MA completed the data analysis. TC and ADF supervised the project. ADF wrote the manuscript with contribution from all authors.

**Disclosure.** The authors declare that they have no conflict of interest.

**Supplemental document.** See Supplement 1 for supporting content.

**Data availability.** The data underpinning this work can be accessed at [DOIdata]